\documentclass[aps, twocolumn, prd, preprintnumbers, amsmath, amssymb, amsfonts, superscriptaddress, nofootinbib]{revtex4-1}
\pdfoutput=1

\usepackage[pagebackref=false,hidelinks]{hyperref}

\usepackage{setspace,latexsym}
\usepackage{color}
\usepackage{epsfig}
\usepackage{graphicx}
\usepackage{slashed}
\usepackage[export]{adjustbox}

\newcommand{\beq}{\begin{equation}}
\newcommand{\eeq}{\end{equation}}
\newcommand{\bea}{\begin{eqnarray}}
\newcommand{\eea}{\end{eqnarray}}
\newcommand{\nn}{\nonumber}

\newcommand{\meV}{\mathrm{meV}}
\newcommand{\eV}{\mathrm{eV}}
\newcommand{\keV}{\mathrm{keV}}
\newcommand{\GeV}{\mathrm{GeV}}
\newcommand{\MeV}{\mathrm{MeV}}
\newcommand{\TeV}{\mathrm{TeV}}

\newcommand{\eq}{\mathrm{(eq)}}

\newcommand{\hmu}{\hat \mu}

\newcommand\lsim{\mathrel{\rlap{\lower4pt\hbox{\hskip1pt$\sim$}}
    \raise1pt\hbox{$<$}}}
\newcommand\gsim{\mathrel{\rlap{\lower4pt\hbox{\hskip1pt$\sim$}}
    \raise1pt\hbox{$>$}}}

\def\bal#1\eal{\begin{align}#1\end{align}}

\begin{document} 

\title{Leptogenesis driven by majoron}

\author{Eung Jin Chun}
\email{ejchun@kias.re.kr}
\affiliation{Korea Institute for Advanced Study, Seoul 02455, South Korea}

\author{Tae Hyun Jung}
\email{thjung0720@gmail.com}
\affiliation{Particle Theory  and Cosmology Group, Center for Theoretical Physics of the Universe,
Institute for Basic Science (IBS),
 Daejeon, 34126, Korea}

\preprint{CTPU-PTC-23-48}

\begin{abstract}
We propose a leptogenesis scenario where baryon asymmetry generation is assisted by the kinetic motion of the majoron, $J$, in the process of lepton-number violating inverse decays of a right-handed neutrino, $N$.
We investigate two distinct scenarios depending on the sources of majoron kinetic motion: 1) the misalignment mechanism, and 2) the kinetic misalignment mechanism.
The former case can naturally generate the observed baryon asymmetry for the majoron mass $m_J \gsim \,\TeV$ and the right-handed neutrino's mass $M_{N} \gsim 10^{11}\,\GeV$. However, an additional decay channel of the majoron is required to avoid the overclosure problem of the majoron oscillation.
The later scenario works successfully for  $m_J \lsim 100\,\keV$, and  $M_{N} \lsim 10^9\,\GeV$ while $M_N$ can be even far below the temperature of the electroweak phase transition as long as sufficiently large kinetic misalignment is provided.
We also find that a sub-$100\,\keV$ majoron is a viable candidate for dark matter.
\end{abstract}

\maketitle
\section{Introduction}
The seesaw mechanism stands out as one of the most compelling frameworks explaining the lightness of left-handed neutrinos through the heaviness of right-handed neutrinos\,\cite{Minkowski:1977sc, Yanagida:1979as, Gell-Mann:1979vob, Glashow:1979nm,Mohapatra:1980yp, Shrock:1980ct,Schechter:1980gr}.
The strength of the seesaw mechanism lies in the natural realization of the baryon asymmetry of the universe through thermal leptogenesis \cite{Fukugita:1986hr} (see, e.g. Ref.\,\cite{Davidson:2008bu} for a review).
In this scenario, the CP asymmetric decay of right-handed neutrinos generates lepton asymmetry which is transferred into the baryon asymmetry via the weak sphaleron process.  However, the amount of the CP asymmetry is naturally proportional to the mass of the decaying particle leading to the so-called Davidson-Ibarra bound: $M_{N}\gsim 10^9\,\GeV$\,\cite{Davidson:2002qv}.

As the Majorana mass of neutrinos breaks the $B-L$ number which is an anomaly-free accidental symmetry in the standard model (SM), an intriguing question is whether $U(1)_{B-L}$ symmetry breaking is spontaneous or explicit.
If it is broken spontaneously (which is what we assume in this paper), 
the heavy right-handed neutrino mass is a consequence of spontaneously broken $U(1)_{B-L}$ symmetry which accompanies a pseudo-Goldstone boson called the majoron\,\cite{Chikashige:1980ui,Gelmini:1980re}.

In this work, we propose a scenario where a kinetic motion of the majoron, denoted by $\dot\theta$,  provides CP asymmetry in the inverse decay of $N$.
This is a realization of spontaneous baryogenesis \cite{Cohen:1987vi,Cohen:1988kt} in the context of the seesaw mechanism endowed with the majoron.
Our scenario can be further characterized by specifying the origin of $\dot \theta$:
 1) the (conventional) misalignment mechanism~\cite{Preskill:1982cy,Abbott:1982af,Dine:1982ah}, and 2) the kinetic misalignment mechanism~\cite{Affleck:1984fy,Co:2019wyp,Co:2019jts}.

A similar setup of our first case (conventional misalignment) has been studied in Ref.\,\cite{Ibe:2015nfa,Domcke:2020kcp} which did not take into account the dynamics coming from $N$, but considered an effective theory with the five-dimensional Weinberg operator assuming sufficiently high seesaw scale.
In this case,
the $B-L$ number is frozen around its decoupling temperature  $T_{W}\simeq 6\times 10^{12}\,\GeV$, and 
if the majoron mass is $O(10^9)\,\GeV$, the majoron oscillation starts around $T_{\rm osc}\simeq T_W$, which leads to a successful leptogenesis.
Unlike the previous works, we include the effects coming from $N$ which generate the $B-L$ number more efficiently compared to the processes involving the Weinberg operator, and consequently, we find how light the majoron can be.

Our second case (kinetic misalignment) has many common features with Refs.\,\cite{Co:2019wyp,Co:2019jts,Domcke:2020kcp,Co:2020xlh,Harigaya:2021txz, Chakraborty:2021fkp,Co:2022aav,Berbig:2023uzs,Chao:2023ojl}
where the final baryon asymmetry is generically determined at the decoupling temperature of the weak sphaleron process or a $B-L$ number-changing process (in our case, it is around $M_N$).
Variations of axiogenesis augmented by the Weinberg operator have also been suggested in Refs.~\cite{Co:2020jtv,Kawamura:2021xpu,Co:2021qgl}.
In this work, we not only take into account the dynamics of $N$, but also include completely different phenomenology that comes from the majoron property.

\section{Basic features}

The seesaw Lagrangian extended with a global $U(1)_{B-L}$ symmetry is written as
\bal
-\mathcal{L}_{\text{int}} = &\frac{1}{2} \sum_{I} 
y_{N_I} \Phi \bar{N}_{I}^c N_{I} + \sum_{\alpha,\, I} Y_{N,\alpha I} \bar{l}_{\alpha} \tilde H N_{I} +  h.c.,
\label{Eq:Lagrangian}
\eal
where $\Phi$ is a complex scalar field with the $B-L$ charge $+2$,  $N_I$ are the right-handed neutrinos with $I=1,2,3$, $l_\alpha$ are the left-handed lepton doublets with $\alpha=e,\mu,\tau$, and $\tilde H\equiv i\sigma_2 H^*$ is the Higgs doublet coupling to up-type quarks and RHNs.
After the $B-L$ breaking, $\Phi$ is replaced by
\bal
\Phi \to \frac{f_J}{\sqrt{2}} e^{i J/f_J},
\eal
 where $J$ is the majoron field.
We assume that the reheating temperature after the inflationary epoch is lower than the $B-L$ phase transition temperature and
the radial mode of $\Phi$ does not affect the physics we discuss in the following.
However, if the reheating temperature is sufficiently high, the universe undergoes the $B-L$ phase transition which may be first-order and the radial mode can play a crucial role in the context of leptogenesis\,\cite{Huang:2022vkf,Dasgupta:2022isg,Chun:2023ezg}.

Going to the field basis by redefining all the fermionic fields $\psi \to e^{i(B-L)_\psi \, \theta /2}\psi$ where $\theta\equiv J/f_J$ and  $(B-L)_\psi$ denotes the $B-L$ number of $\psi$ (e.g., $(B-L)_{N_I}=-1$),  
removed is the $\theta$ dependence in all the Yukawa and scalar potential terms, and there remains only the derivative coupling of the majoron: $-\partial_\mu \theta \, J^\mu_{B-L}/2$ since $B-L$ is anomaly-free.
In a nonzero $\dot \theta\equiv d\theta/dt$ background, a perturbation in the Hamiltonian density, $\dot \theta \, n_{B-L}/2$, is generated to act as an external chemical potential. Thus,
the source term of $B-L$ asymmetry in the Boltzmann equation, proportional to $\dot \theta$, is generated in every term violating the $B-L$ number. This is the origin of the CP violation required for our leptogenesis.

Unlike the conventional thermal leptogenesis, our scenario generates the lepton asymmetry via the so-called ``wash-out" term which acts to ``wash-in" the CP asymmetry provided by the velocity of the majoron field $\dot \theta$.
Assuming a mass hierarchy between right-handed neutrinos: $M_{N_1} \ll M_{N_2}, \, M_{N_3}$,
the ``wash-in process" is mainly governed by the lightest one $N_1$ (which is denoted by $N$ in the following).
Then, the evolution of the lepton number asymmetry density $n_{\Delta l}\equiv n_l-n_{\bar l}$  is determined by (see Appendix.\,\ref{App:Boltzmann} for the derivation)
\bal
&  \dot n_{\Delta l_\alpha}
+3H n_{\Delta l_\alpha}
=
-\Gamma_{Y_{N,\alpha}}
    \Bigg( 
    \frac{n_{\Delta l_\alpha}}{n_{l_\alpha}^\eq}
    +\frac{n_{\Delta H}}{n_H^\eq}
    - \frac{\dot \theta}{T}   \Bigg) + \cdots
,
\label{Eq:Boltzmann_nL}
\eal
where $n_{\Delta H}= n_H - n_{\bar H}$, $n_{X}^\eq$ is the equilibrium number density of $X$, and the interaction rate $\Gamma_{Y_{N,\alpha}}$ controlled by the neutrino Yukawa coupling $Y_N$ is
\bal
\Gamma_{Y_{N,\alpha}}&=n_N^\eq \frac{K_1(z)}{K_2(z)}
\Gamma_{N\to l_\alpha H},
\eal
with $z=M_N/T$, $\Gamma_{N\to l_\alpha H} \simeq |Y_{N,\alpha 1}|^2 M_{N}/ 16\pi$ (assuming $m_N\gg m_{l_\alpha}, m_H$) and $K_{1,2}$ being the modified Bessel functions. 
We neglect the scattering processes of $\Delta L=1$ such as $NQ_3 \leftrightarrow Lt$ since the effect of the scattering is subdominant to the inverse decay term as in the conventional thermal leptogenesis.

Note that the interaction involved in Eq.\,\eqref{Eq:Boltzmann_nL} is the inverse decay, and we do not have a decay term at the tree level.
One may wonder about the effect coming from the helicity asymmetry $n_{\Delta N} = n_{N_+}-n_{N_-}$ where $N_+$ and $N_-$ denote $N$ with positive and negative helicity, respectively.
If the decay term with $n_{\Delta N}$ existed,
it would cancel the $\dot \theta$ contribution since $n_{\Delta N}$ is also shifted proportionally to $\dot \theta$, and the helicity of $N$ can be identified by the chirality (and thus the lepton number) in the $M_N\to 0$ limit.
However, although the helicity asymmetry is indeed generated proportionally to $\dot \theta$ at high temperature $T>\gg M_N$ (see Appendix.\,\ref{App:dispersion} for detail), we find that $n_{\Delta N}$ dependence does not appear in Eq.\,\eqref{Eq:Boltzmann_nL}
because the decay rate of $N_\pm \to l_\alpha H$ is the same as that of $N_\pm \to \bar l_\alpha \bar H$ independently of $N$'s momentum. 
Remark that we consider the case where the CP-violating decay of $N$ is absent or sufficiently suppressed.

\section{Leptogenesis driven by majoron}

We focus on the inverse decay which ``washes in" the CP asymmetry provided by $\dot\theta$ to the lepton sector.
Then, it is 
transferred to the baryon asymmetry by the electroweak sphaleron.
To maximize the efficiency of the wash-in process, the inverse decay is required to be in thermal equilibrium which happens in the so-called strong wash-out regime satisfying $\Gamma_N > H(T_N)$ with $T_N=M_N$.
Therefore, it is important to determine the temperature range at which the weak sphaleron rate and wash-in rate exceed the Hubble expansion rate.

For the weak sphaleron rate, there is a suppression factor of $\exp(-E_{\rm sph}/T)$ where $E_{\rm sph}$ is the energy of the sphaleron configuration that rapidly increases in the broken phase proportionally to the Higgs vev $\langle h \rangle$  at $T$\,\cite{Kuzmin:1985mm,DOnofrio:2014rug}.
Therefore, it gets highly suppressed after the electroweak phase transition, so we consider it to be turned off at $T< T_{\rm EW} \simeq 130\,\GeV$.
On the other hand, when $\langle h \rangle =0$ at high temperature, the sphaleron rate is approximately given by $\alpha_W^5 T$, and it gets decoupled at $T>T_{\rm ws} \simeq 2.5 \times 10^{12}\,\GeV $.

Since we use the wash-in term to generate lepton asymmetry, we \emph{have to} be in the strong wash-out regime; $\Gamma_N \gsim H(T_{N})$.
Taking the usual parameter of the effective neutrino mass
\bal
\tilde m_\nu \equiv \sum_\alpha |Y_{N,\alpha 1}|^2 \frac{v_h^2}{2M_{N}},
\label{Eq:Ya1}
\eal
the strong wash-out condition is $K\equiv \tilde m_\nu/\meV>1$.
For the atmospheric neutrino mass scale of  $\tilde m_\nu =0.05\,\eV$ ($K=50$) and $|Y_{Y, 11}|^2\simeq |Y_{Y, 21}|^2 \simeq |Y_{Y, 31}|^2$,  the inverse decay rate is active when 
\bal
&H< \gamma^{\rm ID}_{Y_N,\alpha}
=
\frac{n_N^\eq}{n_{l_\alpha}^\eq}
\frac{K_1(z)}{K_2(z)}
\Gamma_{N\to l_\alpha H} 
\nn
\\
&\Leftrightarrow \qquad
M_{N}/z_{\rm fo} \lsim T \lsim M_{N}/z_{\rm in}
\eal 
where $z_{\rm in} \simeq 0.7$ and $z_{\rm fo} \simeq 10$.
In order to see the parametric dependence, we keep $z_{\rm in}$ and $z_{\rm fo}$ unless we numerically evaluate.  Then, the baryon asymmetry generation assisted by the majoron is determined at $T_{B-L} = M_{N}/z_{\rm fo}$.

When the weak sphaleron and the wash-in processes are strong enough\,\footnote{
Here, a ``strong enough" reaction means not only to have a reaction rate greater than the Hubble rate but also to be greater than the inverse time scale of changing $\dot \theta$, $(\Delta t)^{-1}_{\dot \theta} \simeq \ddot \theta/\dot \theta$.
}, the baryon number settles down to the equilibrium value which we parameterize as
\bal
n_{B}^{\rm (eq)} = \frac{c_{B} }{6} \dot \theta T^2 , 
\label{Eq:nB_eq}
\\
n_{L}^{\rm (eq)} = \frac{c_{L} }{6} \dot \theta T^2 ,
\label{Eq:nL_eq}
\eal
where $n_{B,\, L}$ are the number density of the baryon and lepton numbers accounting only for SM fermions.
$c_{B}$, $c_{L}$ and $c_{B-L}=c_B-c_L$ for different temperature range are summarized in appendix.\,\ref{App:Equilibrium}.

\subsection{(Conventional) Misalignment mechanism}

Let us first consider the initial condition of $\dot \theta_0 = 0$ and $\theta_0\neq 0$ at the high temperature $T_0$ (which should not be greater than the critical temperature $\sim f_J$ above which the $U(1)_{B-L}$ symmetry is restored).  The classical amplitude $\theta$ starts coherent oscillation when the Hubble rate becomes comparable to its mass $m_J$.
The equation of motion can be written as
\bal
&\ddot \theta + 3H \dot \theta = -\frac{1}{f_J^2} V'(\theta)\simeq -m_J^2  \sin (\theta) 
\eal
where we assumed $V(\theta) = m_J^2 f_J^2 (1-\cos \theta)$ comes from an explicit breaking term of $B-L$ symmetry in the potential: $V(\Phi)=\Phi^{n+4}/\Lambda^n+h.c.$.
When the initial misalignment angle of $\theta$ is not close to $\pi$, one can approximate $V(\theta)\simeq \frac{1}{2}m_J^2 f_J^2 \theta^2$, and obtain 
\bal
\theta(t) \simeq \theta_0 \Gamma(5/4) \left( \frac{2}{m_J t} \right)^{1/4}
J_{1/4} (m_J t),
\label{Eq:oscillation}
\eal
in the radiation-dominated universe ($H \simeq 1/2t$). Here, $J_\alpha$ is the Bessel function of the first kind.

The behavior of Eq.\,\eqref{Eq:oscillation} can be understood separately before and after the oscillation temperature, $T_{\rm osc}$, which is defined by $3H(T_{\rm osc})=m_J$;
\bal
T_{\rm osc} = 5\times 10^8\,\GeV 
\left( \frac{g_{*}}{100} \right)^{-1/4}
\left( \frac{m_J}{\GeV} \right)^{1/2},
\label{Eq:Tosc}
\eal
where $g_{*}$ is the effective number of relativistic degrees of freedom.
By using $J_{1/4}(x) \simeq \Gamma(5/4)^{-1} (x/2)^{1/4} (1 -x^2/5)$ for $x\ll 1$, and $J_{1/4}(x) \sim (2/\pi x)^{1/2} \cos(x-3\pi/8)$ for $x\gg 1$, we obtain an approximate form of
\bal
\dot \theta(T) \sim 
\begin{cases}
\theta_0  m_J  \left( \frac{T_{\rm osc}}{T} \right)^2 
&
\text{for $T>T_{\rm osc}$,}
\\
\theta_0  m_J \left( \frac{T}{T_{\rm osc}}\right)^{3/2} \cos(m_J t)
&
\text{for $T<T_{\rm osc}$,}
\end{cases}
\eal
where we used $H\simeq \frac{m_J}{3} (\frac{T}{T_{\rm osc}})^2$ and neglected order one factors (including signs) and phase shift.
As $T$ decreases from a high temperature, $\dot \theta$ increases and gets maximized around $T_{\rm osc}$.
Then, it starts oscillation with its amplitude being red-shifted as $(T/T_{\rm osc})^{3/2}$.
Therefore, one can expect that the baryon asymmetry generation will be maximized when $T_{\rm osc}$ coincides with $T_{\rm B-L}$.

\begin{figure}[t] 
\begin{center}
\includegraphics[width=0.45\textwidth]{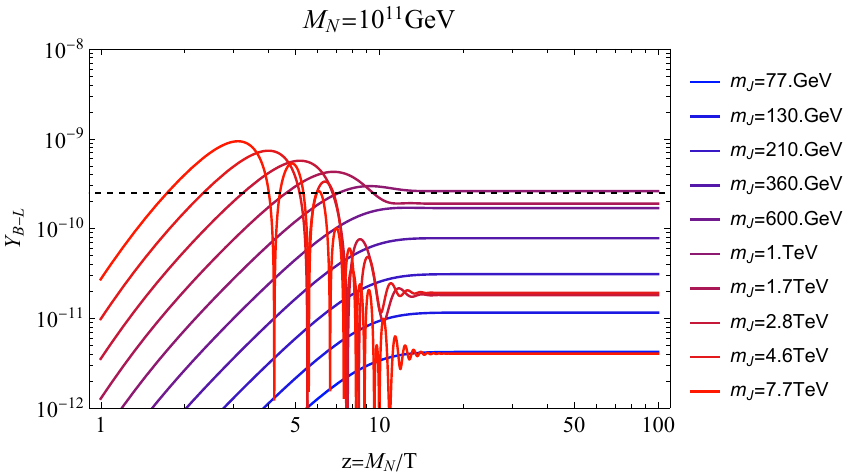} 
\includegraphics[width=0.45\textwidth]{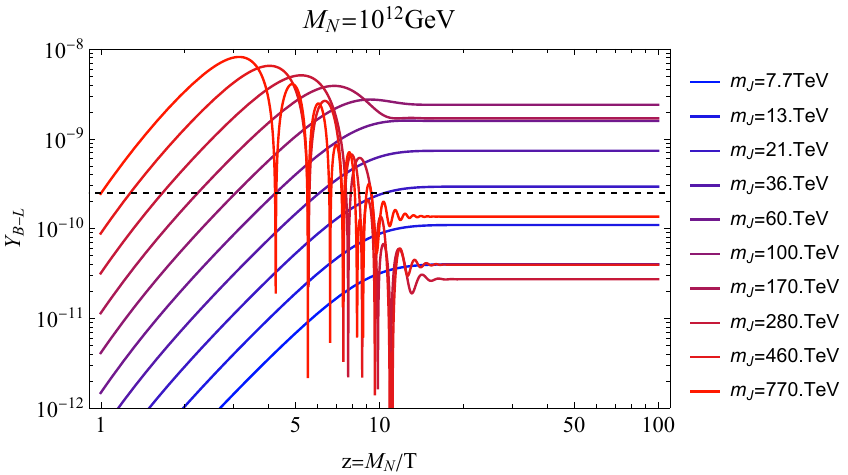} 
\end{center}
\caption{Evolutions of $Y_{B-L}$ for different values of $m_J$ and $M_{N}=10^{11}\,\GeV$ (upper panel) and $10^{12}\,\GeV$ (lower panel). For a fixed $M_{N}$, $Y_{B-L}$ is maximized when the parameters satisfy $T_{\rm osc} \simeq T_{B-L}$.
The horizontal dashed line corresponds to the $Y_{B-L}$ value required for the observed baryon asymmetry.}
\label{Fig:osc_evolution}
\end{figure}

To investigate further detail, let us, first, consider the case when $T_{\rm osc} < T_{B-L} $.
Since the weak sphaleron and the wash-in rates are strong enough for $T>T_{B-L}$, $n_{B-L}$ follows the equilibrium values \eqref{Eq:nB_eq} and \eqref{Eq:nL_eq} adiabatically, and gets frozen before the oscillation starts.
Therefore, we can estimate
\bal
Y_{B-L} &\simeq \left. \frac{1}{s(T)}\frac{1}{6} c_{B-L} \dot \theta(T)T^2 \right|_{T=T_{B-L}} \nonumber
\\
&\sim \frac{c_{B-L} \theta_0}{g_*(T_{B-L})}\frac{m_J}{T_{B-L}} \left( \frac{T_{\rm osc}}{T_{B-L}} \right)^2 ,
\label{Eq:YBL_osc_1}
\eal
where $s(T)=\frac{2\pi^2}{45}g_* T^3$ is the total entropy density of the background plasma with the effective number of relativistic degrees of freedom $g_*$.
As $T_{\rm osc} \propto m_J^{1/2}$ (see Eq.\,\eqref{Eq:Tosc}), we have the proportionality of $Y_{B-L} \propto m_J^2$ for $T_{\rm osc} < T_{B-L}$ when $T_{B-L}$ is fixed.

On the other hand, when $T_{\rm osc} > T_{B-L}$, the oscillation starts first.
Since the oscillation time scale, $\Delta t_{\rm osc} \simeq m_J^{-1}$ becomes shorter than the Hubble time scale,
it is not guaranteed for the $B-L$ number to settle down at the equilibrium value.
Assuming that $m_J> \Gamma_{Y_{N,\alpha}}>H$, $Y_{B-L}$ can be estimated as
\bal
Y_{B-L}(T) &\sim \frac{c_{B-L} \Gamma_{Y_{N,\alpha}} }{T s(T)} \int dt \, \dot \theta \nonumber
\\
&\sim
\frac{c_{B-L} \theta_0}{g_*(T)}
\frac{\Gamma_{Y_{N,\alpha}} }{T^4} 
\left( \frac{T}{T_{\rm osc}} \right)^{3/2}
\sin(m_J t).
\eal
At the temperature around $T_{B-L}$, $Y_{B-L}$ is frozen during the oscillation.
Taking the approximation of $\Gamma_{Y_{N,\alpha}}\simeq H(T_{B-L}) T_{B-L}^3$, we obtain
\bal
|Y_{B-L}|&\lsim  \frac{c_{B-L} \theta_0}{g_*(T_{B-L})}  \frac{H(T_{B-L})}{T_{B-L}}\left( \frac{T_{B-L}}{T_{\rm osc}}\right)^{3/2}  
\label{Eq:YBL_osc_2}
\eal
which shows the proportionality of $Y_{B-L} \propto m_J^{-3/4}$ for a fixed $T_{B-L}$.

These features can be seen in Fig.\,\ref{Fig:osc_evolution} where we depict the evolution of $Y_{B-L}$ as a function of  $z=M_{N}/T$ by solving the full set of the Boltzmann equations summarized in Appendix.\,\ref{App:Boltzmann}.
Considering two different values of $M_{N}=10^{11}\,\GeV$ (upper panel) and $10^{12}\,\GeV$ (lower panel), we show the dependence of $Y_{B-L}$ on the values of $m_J$ which is scanned around $T_{\rm osc} \sim T_{\rm B-L}$.
As we discussed previously, the frozen value of $Y_{B-L}$ is maximized when $T_{\rm osc} \simeq T_{\rm B-L}$.

\begin{figure}[t] 
\begin{center}
\includegraphics[width=0.45\textwidth]{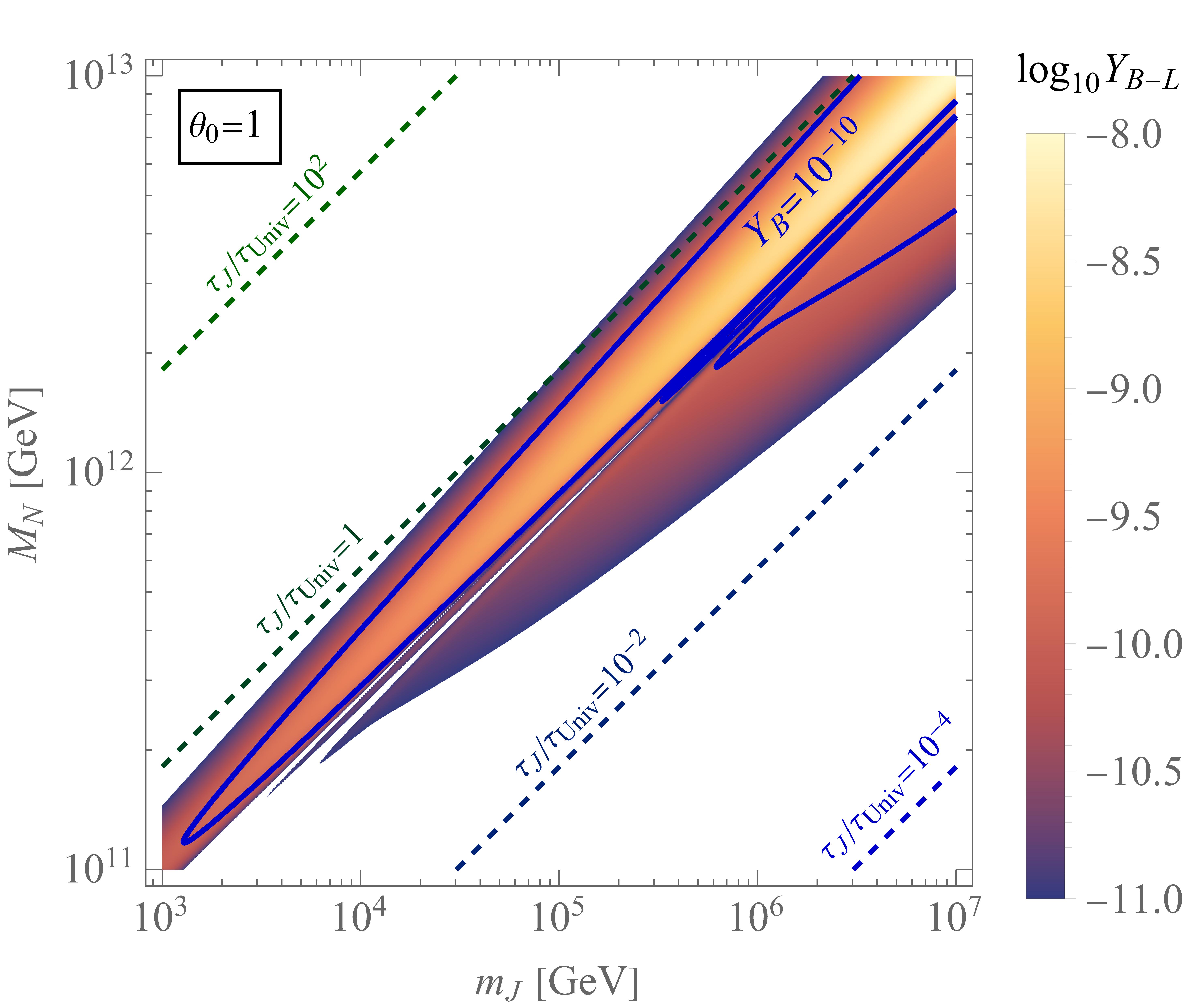} 
\end{center}
\caption{The expected value of $Y_B$ is depicted as a function of $m_J$ and $M_{N}$ for $\theta_0=1$.
Dashed lines show the lifetime of majoron in the minimal setup.}
\label{Fig:osc_scan}
\end{figure}

From the previous estimations, we conclude that $Y_{B-L}$ is bounded from above for a fixed $T_{B-L}$, and the maximized value at $T_{B-L}\simeq T_{\rm osc}$ is given by
\bal
Y_{B-L}^{\rm max}(T_{B-L})\sim 10^{-9} \,
c_{B-L} \theta_0 
\left(
    \frac{100}{g_*(T_{B-L})}
\right)
\left(
    \frac{T_{B-L}}{10^{10}\,\GeV}
\right),
\eal
which we obtain from Eq.\,\eqref{Eq:YBL_osc_1} or Eq.\,\eqref{Eq:YBL_osc_2} taking $T_{\rm osc} \simeq T_{B-L}$ (and Eq.\,\eqref{Eq:Tosc} to remove $m_J$ dependence), and including $O(10)$ factor that arises from our numerical solution of the Boltzmann equations.
This implies that, for $Y_B \simeq \frac{28}{79} Y_{B-L} \simeq 8.7\times 10^{-11}$, we need
\bal
T_{B-L} \gsim 10^{10}\,\GeV 
\quad \Rightarrow \quad
M_{N} \gsim 10^{11}\,\GeV \left( \frac{z_{\rm fo}}{10}\right),
\label{Eq:condition}
\eal
considering $\theta_0=O(1)$.
For the rigorous results, we solve the full Boltzmann equations,  and show the final value of $Y_B$ in the plane of $m_J$ and $M_{N}$ in Fig.\,\ref{Fig:osc_scan} taking $z_{\rm fo}=10$ and $\theta_0=1$. 
In the plot, we also show the lifetime of the majoron which is determined by its dominant decay channel $J\to \nu \nu,\, \bar \nu \bar \nu$, and thus has the decay rate proportional to $m_\nu^2/f_J^2$. One can see that the majorons are fairly long-lived in the parameter region of our interest.  This causes a serious problem of overclosing the universe.

The energy density of the majoron oscillation is indeed given by
\bal
\hspace{-0.5cm}
&\frac{\rho_{\rm osc}(T_{\rm osc})}{s(T_{\rm osc})} 
\simeq \frac{\theta_0^2 m_J^2 f_J^2}{s(T_{\rm osc})}
\\
&\sim 
0.4\,\eV \, 
\theta_0^2 \,
\left( \frac{10}{z_{\rm fo}}\right)
\left( \frac{0.1}{y_N}\right)^2
\left( \frac{T_{\rm osc}}{T_{B-L}}\right)
\left( \frac{M_{N}}{4\times 10^8\,\GeV} \right)^{3} , \nonumber
\eal
 which is unacceptably large for $M_N> 10^{11} {\rm GeV}$.

To circumvent this problem, one may introduce an additional coupling of the majoron, such as $(e^2/32\pi^2)(J/f_J)  F^{\mu\nu}\tilde F_{\mu\nu}$ whose UV completion can be done by adding vector-like charged leptons (while the $B-L$ charge of vector-like leptons should be assigned chirally).
This can drastically increase the decay rate to make the majoron decay away to two photons before Big Bang nucleosynthesis (BBN) but after the leptogenesis era.
Introducing this operator does not change the previous estimation of the leptogenesis part\,\footnote{
The $c_{J\gamma}(J/f_J)F^{\mu\nu}F_{\mu\nu}$ interaction can, in principle, generate friction to the $\theta$ motion via the tachyonic instability of photons.
However, this effect is small in the case of conventional misalignment scenarios because the wavelength of the tachyonic mode is always greater than the Hubble radius unless the coefficient $c_{J\gamma}$ is greater than order one.
}.

\subsection{Kinetic misalignment mechanism}
Now, let us consider the case when $\dot \theta \neq 0$ at $T\gg T_{\rm osc}$.
This can be realized by the so-called kinetic misalignment mechanism\,\cite{Affleck:1984fy,Co:2019wyp,Co:2019jts}.
Assuming a sufficiently flat potential of $B-L$ breaking field $\Phi$, its radial mode $\phi$ can be stuck at a large field value due to the Hubble friction.
When the Hubble rate becomes comparable to the curvature of the potential, $\phi$ starts rolling down, and an explicit breaking of $B-L$ symmetry, which also generates $m_J$, drives the motion along the majoron direction.

We treat the initial majoron motion as a free parameter since it strongly depends on the potential shape of $\Phi$.
Therefore, our starting point is taking nonzero $\dot \theta (T_0) = \dot \theta_0$ at a sufficiently high temperature $T_0$ (but still much lower than $f_J$ to avoid thermal friction).
As in Ref.\,\cite{Co:2020xlh}, we take a free parameter $Y_\theta \equiv f_J^2 \dot \theta_0/s(T_0)$ which is approximately conserved throughout the leptogenesis process, i.e. $f_J^2 \dot \theta (T)/s(T) \simeq Y_\theta$ or $\dot \theta(T) \simeq Y_\theta s(T)/f_J^2$.

If $T_{\rm EW}< T_{B-L} $, the $B-L$ number is frozen before the electroweak phase transition.
Then, the $B-L$ number is re-distributed, and the baryon number is finally frozen at $T_{\rm EW}$ as
\bal
Y_B = \frac{28}{79} Y_{B-L}^{\rm (eq)}(T_{B-L})
= \frac{14}{237} c_{B-L} Y_\theta \left( \frac{y_N}{\sqrt{2} z_{\rm fo}} \right)^2,
\label{Eq:YB_kin_1}
\eal
where we took the replacement: $\dot \theta (T)= Y_\theta \, s(T)/f_J^2$.

On the other hand, if $T_{\rm EW}> T_{B-L} $, the baryon number is frozen at the electroweak phase transition during the $B-L$ number is changing.
Therefore, the baryon asymmetry is given by the equilibrium value at $T_{\rm EW}$:
\bal
Y_B = Y_{B}^{\rm (eq)}(T_{\rm EW}) = 
\frac{1}{6} c_{B} Y_\theta \left( \frac{T_{\rm EW}}{f_J} \right)^2  
.
\label{Eq:YB_kin_2}
\eal
which is valid only for 
$M_{N} \gsim  z_{\rm in}\,  T_{\rm EW}$.
In this case, the decay processes such as $N\leftrightarrow l H$ or $N\leftrightarrow \bar l \bar H$ may be prohibited if $M_N$ is lighter than the Higgs mass (including thermal corrections).
Instead, $H\leftrightarrow l N$ and $\bar H\leftrightarrow \bar l N$ become responsible for the main $B-L$ number-changing process.
Nevertheless, 
due to the dependence of $z_{\rm in}\approx (K/2)^{-1/3}$ for large $K$, $z_{\rm in}$ is insensitive to the detailed dependences, and 
the validity of Eq.\,\eqref{Eq:YB_kin_2} requires $M_N$ greater than $O(10)\,\GeV$ even when we allow a tuning in the $Y_{N,\alpha 1}$ structure.

When $M_N \lsim z_{\rm in} T_{\rm EW}$, there exists an additional suppression factor of $\gamma^{\rm ID}(T_{\rm EW})/H(T_{\rm EW})$;
\bal
Y_B \sim  
\frac{1}{6} c_{B} Y_\theta \left( \frac{T_{\rm EW}}{f_J} \right)^2 
\left( \frac{\gamma^{\rm ID}(T_{\rm EW})}{H(T_{\rm EW})}\right),
\eal
where
\bal
\frac{\gamma^{\rm ID}(T_{\rm EW})}{H(T_{\rm EW})}
\simeq
 10
\left(
    \frac{K}{50}
\right)
\left(
    \frac{g_*}{100}
\right)^{-1/2}
\left(
    \frac{M_N}{T_{\rm EW}}
\right)^3
\label{Eq:YB_kin_3}
\eal
with $\gamma^{\rm ID}=\sum_\alpha \gamma^{\rm ID}_{Y_N,\alpha}$.
Therefore, $Y_\theta$ needs to be even greater to compensate for this suppression factor.

For the validity of our consideration, the kinetic energy density of the majoron needs to be smaller than the radiation energy density at least when the $B-L$ number or $B$ number is frozen\,\footnote{
Our mechanism may work even during the kination domination with an appropriate change of the Hubble rate, which needs a further scrutiny. In this article, we limit ourselves to the radiation domination which does not require too small $y_N$.
}.
This implies $f_J^2 \dot\theta(T_*)^2/2 < \pi^2 g_*(T_*) T_*^4/30$, where $T_* = \max(T_{B-L}, T_{\rm EW})$.
For $T_{B-L}>T_{\rm EW}$, using the condition \eqref{Eq:YB_kin_1}, we obtain
\bal
y_N \gsim 
10^{-7}
\left( 
    \frac{g_*}{100}
\right)^{1/2}
\left( 
    \frac{z_{\rm fo}}{10}
\right)
\left( 
    \frac{1}{c_{B-L}}
\right),
\label{Eq:yN_lower_bound}
\eal
where we take $Y_B\simeq 8.7\times 10^{-11}$ for the observed baryon asymmetry\,\cite{Planck:2018vyg}.
Similarly, by using \eqref{Eq:YB_kin_3} when $T_{B-L} < T_{\rm EW}$, we obtain the lower bound of $M_N$
\bal
M_N > 2\,\GeV
\left(
    \frac{f_J}{10^6\,\GeV}
\right)^{1/3}
\left(
    \frac{g_*}{100}
\right)^{1/3}
\left(
    \frac{1}{c_B}
\right)^{1/3}
\left(
    \frac{50}{K}
\right)^{1/3}.
\eal
As we will show below, $f_J$ needs to be greater than $10^6\,\GeV$ to avoid the constraint from CMB and BAO analysis\,\cite{Audren:2014bca,Enqvist:2019tsa,Nygaard:2020sow,Alvi:2022aam,Simon:2022ftd}, so this puts the lower bound $M_N \gsim 2\,\GeV$.

On the other hand, if the temperature $T_0$ when $\dot \theta$ is initially generated is large compared to $T_*$,
there exists a temporary kination domination (KD) era during which the kinetic energy of $\theta$ dominates the universe.
Although this would not change our baryogenesis analysis, there can be a significant enhancement of gravitational waves during the transition to KD from radiation domination (RD) or matter domination (MD) after the inflation and vice versa\,\cite{Co:2021lkc,Gouttenoire:2021wzu,Gouttenoire:2021jhk,Harigaya:2023mhl}.

It is also remarkable that the initial kinetic misalignment required for successful leptogenesis can generate the right amount of dark matter abundance from the coherent oscillation of the majoron occurring at a later time.  
As $\dot \theta$ gets redshifted as $s(T)\sim T^3$, the kinetic energy density of majoron scales as $f_J^2 \dot \theta^2/2\propto T^6$, and becomes eventually comparable to the potential barrier $m_J^2 f_J^2$.
Once it happens, the majoron gets trapped in the potential.
The trapping temperature can be estimated by $f_J^2 \dot \theta^2/2 \simeq m_J^2 f_J^2$ leading to the relation
\bal
s(T_{\rm trap}) \simeq \frac{m_J f_J^2}{Y_\theta}.
\eal
Then, the trapped majoron can either 1) start oscillation immediately ($m_J>3H(T_{\rm trap})$, i.e.  $T_{\rm osc}>T_{\rm trap}$), or 2) start oscillation after a while ($m_J<3H(T_{\rm trap})$, i.e. $T_{\rm osc}<T_{\rm trap}$).
For the first case, the oscillation energy density is frozen as
\bal
\frac{\rho_{\rm osc}}{s} \sim \frac{m_J^2 f_J^2}{s(T_{\rm trap})} \sim m_J Y_\theta.
\label{Eq:rho_osc_1}
\eal
On the other hand, if $T_{\rm osc}<T_{\rm trap}$, the majoron is stuck at an $O(1)$ intermediate value $\theta =\theta_0$, and starts oscillation at $T_{\rm osc}$.
The abundance in this case is given by
\bal
\frac{\rho_{\rm osc}}{s} \simeq \frac{\theta_0^2 m_J^2 f_J^2}{s(T_{\rm osc})},
\label{Eq:rho_osc_2}
\eal
from which one finds a fixed relation between $m_J$ and $f_J$ to explain the observed dark matter abundance.

\begin{figure}[t] 
\begin{center}
\includegraphics[width=0.45\textwidth]{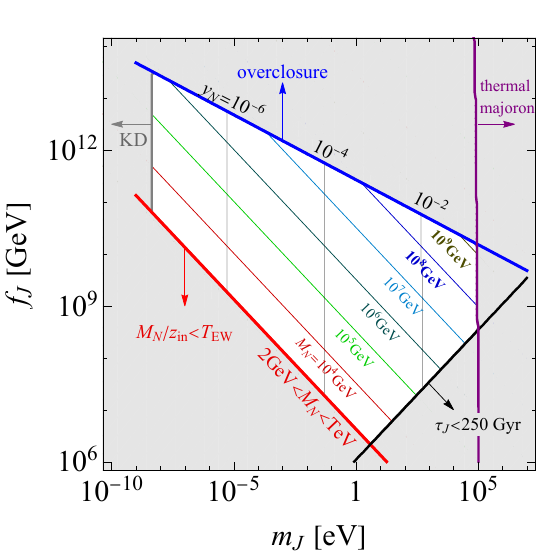} 
\end{center}
\caption{Available parameter space for the kinetic misalignment case.
The gray region is excluded by the overclosure of the universe (above the blue line), by the condition for the leptogenesis (below the red line), by the thermal relic of majorons (right to the purple line), by the kination domination (left to the gray line), or by the constraints from CMB and BAO (below the black line).
Colored lines and vertical gray lines inside the allowed region (white) show the required values of $M_{N}$ and $y_N$, respectively. 
}
\label{Fig:model_parameter_space}
\end{figure}

In Fig.\,\ref{Fig:model_parameter_space}, we show the parameter space (white) that is consistent with the observed baryon asymmetry and dark matter density at present\,\cite{Planck:2018vyg}:
\bal
Y_B \simeq 8.7\times 10^{-11}, \quad \frac{\rho_J}{s}\simeq 0.44\,\eV.
\eal
At each point in the $(m_J, f_J)$ plane, these two conditions fix the parameters $Y_\theta$, and $y_N$ (gray lines) or $M_{N}$ (colored lines).
Here, we take $z_{\rm fo}\simeq 10$.
For the consistency of the scenario, we require the following conditions:
\begin{itemize}
\item[1.] The lower limit of the majoron lifetime {$\tau_J > 250\,{\rm Gyr}$ to avoid the constraint from CMB and BAO analysis \cite{Audren:2014bca,Enqvist:2019tsa,Nygaard:2020sow,Alvi:2022aam,Simon:2022ftd} (see the black line). }
\item[2.] When $\Gamma_{HL \leftrightarrow J N} > H$, thermal majorons can be produced efficiently when they are relativistic, and thus their relic energy density may become too large (see Ref.\,\cite{Sabti:2019mhn,Blinov:2019gcj,Sandner:2023ptm,Chang:2024mvg} for corresponding strong constraints when their population becomes large). 
The production rate can be approximated as $\Gamma_{HL \leftrightarrow J N} \approx |Y_{N,\,\alpha 1}|^2 |y_N|^2T/8\pi$ for $T>M_N$ while $\Gamma_{HL \leftrightarrow J N}$ for $T<M_N$ is negligible due to the Boltzmann suppression and the $T^2/f_J^2$ suppression.
Then, the thermalization condition is met for $1<T/M_N \lsim 50\,y_N^2 (K/50)(g_*/100)^{-1/2}$.
Therefore, we demand $y_N \lsim 0.14\,(K/50)^{-1/2}(g_*/100)^{1/4}$ to avoid the overproduction (see the purple line).
\end{itemize}
Consequently, we obtain $m_J<100\,\keV$ (see the purple line), and $f_J>10^6\,\GeV$ (see the black line).

\section{Discussions on phenomenology}
\label{sec:pheno}
Searching for heavy neutral leptons (HNLs) like $N$ is one of the most active research fields, and can test the low $M_N$ region of our second scenario (see \cite{Chun:2019nwi,Abdullahi:2022jlv} and references therein).
For $M_N \approx 0.2-6\,\GeV$, rare meson decays put the bounds like $K/50\lsim 30$ for $M_N< 2\,\GeV$ and $K/50\lsim 10^3$ for $M_N=2-6\,\GeV$\,\cite{Chun:2019nwi}.
The mass range of $M_N = O(10)\,\GeV$, can be tested at future colliders such as FCC-ee and FCC-hh if $K/50 \gsim O(10)$  \cite{Abdullahi:2022jlv}.

On the other hand, a direct test of majoron is very challenging. 
All the majoron couplings to the SM particles involve $Y_N^2$ and thus are generically suppressed by $m_\nu/f_J$ or $m_\nu/v_h$.
This behavior can be seen from the Lagrangian \eqref{Eq:Lagrangian} where $\Phi$ and $N_I$ completely decouple from the SM sector in the limit of $m_\nu \to 0$ corresponding to $Y_{N,\alpha I}\to 0$.
The $m_\nu$-suppression at higher loop order can be explicitly seen in Ref.\,\cite{Heeck:2019guh}.
For instance, majoron to photon-photon coupling can be generated at two-loop order, but it is very challenging to leave an observable signature for small $m_J$ because of the $m_\nu$ suppression (see Appendix.\,\ref{App:photonphoton} for details).

The $m_\nu$-suppression makes it (almost) impossible to test the model except for the high $K$ limit discussed in the beginning of this section.
Although the supernova constraints seem strong in terms of the coupling strength ($g \simeq \tilde m_\nu/f_J \lsim 10^{-10} (m_J/100\,\MeV)^{-1}$ for $100\,\keV \lsim m_J \lsim 100\,\MeV$ and $g\lsim 10^{-7}$ for $m_J \lsim 100\,\keV$\,\cite{Choi:1987sd, Choi:1989hi,Chang:1993yp,Akita:2022etk, Fiorillo:2022cdq,Akita:2023iwq}), the constraint in terms of its decay constant is only $f_J \gsim 10\,\GeV$ at most.
Neutrinoless double beta decay experiments also put constraints on the majoron coupling to $\nu_e$ via searching for majoron-emitting channel as $g_{ee} \sim m_{\nu, ee}/f_J \gsim 10^{-5}$\,\cite{Arnold:2018tmo,Barabash:2018yjq,Kharusi:2021jez,GERDA:2022ffe,CUPID-0:2022yws}.

Various cosmological constraints on the majoron abundance come from the analysis of CMB and BBN\,\cite{Sabti:2019mhn,Blinov:2019gcj,Sandner:2023ptm,Chang:2024mvg}.
However, in our scenarios, the coupling of majoron is so small that majorons are not thermally produced (once we avoid $HL\leftrightarrow JN$ as discussed in the previous section), so majorons neither change the expansion rate nor drive early matter domination.
CMB also puts a constraint on neutrino self-interaction mediated by the majoron even when the majoron abundance is small,  as discussed in Ref.\,\cite{Sandner:2023ptm}, and the corresponding constraint is $g\lsim 10^{-12}$ ($f\gsim 10\,\GeV$) for $m_J \lsim \keV$.

Despite the intrinsic suppression factor in the coupling strength, the majoron \emph{dark matter scenarios} can have some interesting impact in the CMB and BAO observation\,\cite{Audren:2014bca,Enqvist:2019tsa,Nygaard:2020sow,Alvi:2022aam,Simon:2022ftd}. This puts the limit $\tau_J > 250\,{\rm Gyr}$ which was taken into account in the previous section.
Although excluded in our scenario, the majorana mass above MeV is severely constrained by the measurements of neutrino flux\,\cite{Borexino:2019wln,KamLAND:2021gvi,Olivares-DelCampo:2017feq,Palomares-Ruiz:2007egs,Frankiewicz:2016nyr,Super-Kamiokande:2011lwo,Super-Kamiokande:2021jaq,IceCube:2021kuw,IceCube:2023ies,Arguelles:2022nbl,Albert:2016emp} (see Ref.\,\cite{Akita:2023qiz} for the analysis in the majoron parameters).

\section{Summary}
In this work, we have investigated a leptogenesis scenario where the lepton asymmetry is generated via the decay and inverse decay of the lightest right-handed neutrinos under the CPT violation given by a background majoron motion, $\dot \theta$.
To generate nonzero $\dot \theta$, we have considered two scenarios.
One is generating it via the conventional misalignment mechanism, and the other is generating it via the kinetic misalignment mechanism.

For the misalignment scenario, we find that our scenario successfully generates baryon asymmetry if $M_{N}$ is greater than $10^{11}\,\GeV$.
However, the energy density of the majoron oscillation becomes greater than the observed dark matter abundance while its lifetime is the order of the age of the universe.
The simplest way to avoid this problem is introducing an additional interaction such as $J F \tilde F$ to make the lifetime much shorter.

On the other hand, the leptogenesis scenario sourced by the kinetic misalignment can be realized for $1\,\GeV \lsim M_{N}\lsim 10^9\,\GeV$ and $m_J\lsim 100\,\keV$ while the majoron oscillation can be a viable candidate of the dark matter.
Thus, this scenario can be (partially) tested by searching for heavy neutral leptons.
However, the majoron lighter than $100\,\keV$ is hardly testable as we discussed.
\\

\vskip 1em
{\bf Acknowledgement:}
This work was supported by IBS under the project code IBS-R018-D1.

\begin{appendix}

\section{Majorana fermion and an external chemical potential}
\label{App:dispersion}

In the background of the kinetic motion of majoron field $\dot\theta$, the dispersion relation of a Majorana fermion behaves differently from that of a Dirac or Weyl fermion 
due to the Majorana mass term breaking the $U(1)$ symmetry.
The Lagrangian of a Majorana fermion $\psi$ whose mass is generated after the spontaneous breaking of the global $U(1)$ symmetry is 
\begin{align}
{\cal L}_M &= {1\over2} \left( \bar\psi_L i\gamma^\mu \partial_\mu \psi_L + \bar\psi_R i\gamma^\mu \partial_\mu \psi_R \right) \nonumber \\
& -{1\over2}\left(M e^{i\theta} \bar\psi_L \psi_R + M e^{-i\theta} \bar\psi_R \psi_L \right) + \cdots
\end{align} 
where $\psi_R \equiv \psi_L^{\;c}=N$ and $\theta=J/f_J$ in the case of majoron under consideration.  Removing the $\theta$ dependence in the mass term by the field redefinition $\psi_{L,R} \to e^{\pm i \theta/2} \psi_{L,R}$, one can obtain 
\bal
{\cal L}_M &\to {1\over2} \left( \bar\psi_L i\gamma^\mu \partial_\mu \psi_L + \bar\psi_R i\gamma^\mu \partial_\mu \psi_R \right) \nonumber  \\
& -{1\over2}\left(M \bar\psi_L \psi_R + M e^{-i\theta} \bar\psi_R \psi_L \right) \\
&
-\frac{1}{2}\partial_\mu \theta
(\bar \psi_L \gamma^\mu \psi_L
- \bar \psi_R \gamma^\mu \psi_R)
+ \cdots. \nonumber
\eal 
Note that the induced current interaction term is chiral, unlike the case of other SM fermions where a vector current interaction arises under the $\exp[i(B-L)\theta/2]$ rotation.
This is because of the identity $\psi_R=\psi_L^c$.

The free equations of motion for the $u$-spinors in $\psi_{L,R} \sim u_{L,R} e^{-i p\cdot x}$ are given as follows:
\begin{align}
(p_{-} \cdot \bar\sigma)\, u_L = M u_R 
\label{Eq:uL}
\\
(p_+ \cdot \sigma)\, u_R = M u_L
\label{Eq:uR}
\end{align}
where $p_{\mp \mu} = p_\mu \mp \partial_\mu \theta/2$.
Here, we used the chiral representation of $\gamma^\mu$:
\bal
\gamma^\mu = \begin{pmatrix}
    0 & \sigma^\mu \\ \bar\sigma^\mu & 0
\end{pmatrix} 
~~\mbox{with} ~~
\begin{cases}
\begin{matrix}
  \sigma^\mu=(1,+\vec\sigma) \\
  \bar\sigma^\mu=(1,-\vec\sigma)
\end{matrix} 
\end{cases} 
\eal
where $\vec\sigma$ are the Pauli matrices.

From Eq.~\eqref{Eq:uL} and \eqref{Eq:uR}, we find the dispersion relation 
\begin{equation}
    (p_+\cdot \sigma) \, (p_-\cdot\bar\sigma)u_L=M^2
\end{equation}
which leads to two distinct solutions for the helicity eigenstates ${\cal H}= \vec\sigma \cdot \vec p/{\rm p}= \pm 1 $ with ${\rm p} \equiv |\vec p|$.
We do not present solutions for $u_R$ and $v_{L,R}$ (where $\psi_{L,R}\sim v_{L,R}\,e^{ip\cdot x}$) because the degrees of freedom is effectively two, which are identified by the relations with $u_L$;
for instance, $u_R$ is fixed by Eq.\,\eqref{Eq:uL}, and $v_{L,R}$ are fixed by $\psi_R=\psi_L^c$.

For the homogeneous background, $\dot \theta \neq 0$ and $\partial_i \theta=0$, one finds 
\begin{equation}
    E=\sqrt{{\rm p}^2 + M^2 +{1\over4}\dot\theta^2 -  {\cal H} \dot\theta\, {\rm p}}.
\end{equation}
In the limit of $ |\dot\theta|/E_0 \ll 1$ with $E_0\equiv \sqrt{{\rm p}^2+ M^2}$, we have $E\approx E_0 \mp \frac{\dot\theta}{2} \frac{\rm p}{E_0}$ where ${\rm p}/{E_0}$ approaches 1 in the ultra-relativistic (Weyl) limit,
whereas it gets suppressed as ${\rm p}/M_N$ in the non-relativistic limit.

In the Boltzmann approximation, one finds that the equilibrium number density of the Majorana fermion $N$ with the external chemical potential is given by 
\begin{align}
        n_{N_\pm} \approx \frac{T^3}{2\pi^2} \left(z^2 K_2(z) \pm 2 {\dot\theta \over T} e^{-z}(1+z) \right),
        \label{Eq:helicity_asymmetry}
\end{align}
where the $+$ sign ($-$ sign) stands for the positive (negative) helicity.

This can be understood as the $B-L$ conservation in the limit of $M_N\to 0$ since $N$'s helicity is the lepton number.
The scattering processes which does not vanish at $M_N\to 0$, e.g. $NQ_3 \leftrightarrow lt$, are affected by the helicity asymmetry of $N$.

On the other hand, the decay and inverse decay processes are always proportional to $M_N$, and therefore it is not affected by the helicity asymmetry of $N$.
This can be explicitly seen from the fact that the decay rate of $N_{\pm}\to l H$ is the same with $N_{\pm}\to \bar l \bar H$ independently of the inertial frame.

Since, in this paper, we neglect the scattering terms while we only keep the decay and inverse decay terms, Eq.\,\eqref{Eq:helicity_asymmetry} will not be used in our Boltzmann equations.
Note, however, that a more precise estimation including scattering terms should include the helicity asymmetry of $N$, and therefore the equilibrium values of $Y_B$ and $Y_L$ are modified accordingly.

Other SM fermions follow different dispersion
relation as $\psi_L$ and $\psi_R$ are independent degrees (corresponding to particle and anti-particle states, respectively), and they carry the same $U(1)_{B-L}$ charge.
Thus, the modified four-momenta $p_{\mp}$ that appear in Eqs.~\eqref{Eq:uL} and \eqref{Eq:uR} should be replaced by the same sign ones, which gives the dispersion relation of 
\begin{equation}
    E_\psi =\sqrt{{\rm p}^2 + m_\psi^2} \mp  {1 \over2} (B-L)_\psi \dot\theta
\end{equation}
where $m_\psi$ is the Dirac mass.

\section{Boltzmann equations}
\label{App:Boltzmann}

\subsection{Decay and inverse decay of ${N}$}
In this section, we approximate that the distribution function of $X$ is given by $f_X(p)\simeq (n_X/n_X^\eq)f_X^\eq(p)$ with assuming the kinetic equilibrium.
We further approximate $f_X^\eq(p)$ by the Maxwell-Boltzmann distributions for $X=N$, $l_\alpha$, and $H$ for simplicity.
Then, the decay and inverse decay terms of right-handed neutrinos can be written as
\bal
\dot n_{l_\alpha}+ 3H n_{l_\alpha} 
= &+\frac{n_N}{n_N^\eq} \Gamma^\eq(N \to l_\alpha H)
\nn \\
&-\frac{n_l n_H}{n_l^\eq n_H^\eq}
\Gamma^\eq(l_\alpha H\to N) +\cdots
\\
\dot n_{\bar l_\alpha} + 3H n_{\bar l_\alpha} 
=& +\frac{n_{N}}{n_{N}^\eq} \Gamma^\eq(N \to \bar l_\alpha \bar H)
\nn \\
&-\frac{n_{\bar l_\alpha} n_{\bar H}}{n_{l_\alpha}^\eq n_H^\eq}
\Gamma^\eq(\bar l_\alpha \bar H\to  N) +\cdots
\eal
where
\bal
\Gamma_{Y_{N,\alpha}}&\equiv
\Gamma^\eq(N\to l_\alpha H)
=\Gamma^\eq(l_\alpha H \to N) \nonumber
\\
&=
\int \frac{d^3 p_N}{(2\pi)^3} f_N^\eq (p_N)
\frac{M_N}{E_N} \Gamma_{N\to l_\alpha H}
\\
&=n_N^\eq \frac{K_1(z)}{K_2(z)}
\Gamma_{N\to l_\alpha H} \nonumber
\eal
and $\Gamma_{N\to l_\alpha H} = |Y_{N,\alpha 1}|^2 M_N/16\pi$ (one can use $n_N^\eq = \frac{2}{(2\pi)^2}z^2 K_2(z) T^3$ to further simplify the equation).
Note that since $\Gamma(N_+ \to l_\alpha H) = \Gamma(N_- \to l_\alpha H)$ and $\Gamma(N_+ \to \bar l_\alpha \bar H) = \Gamma(N_- \to \bar l_\alpha \bar H)$, the decay terms are combined by $n_N=n_{N_+}+n_{N_-}$.

With nonzero chemical potentials, we can replace $n_{\Delta X}/n_X^\eq \simeq 2\mu_X/T$, where $n_{\Delta X} \equiv n_X - n_{\bar X}$.
Then, the corresponding term in the Boltzmann equation for $n_{\Delta l_\alpha}$ becomes
\bal
&  \dot n_{\Delta l_\alpha}
+3H n_{\Delta l_\alpha}
=
-\Gamma_{Y_{N,\alpha}}
    \Bigg( 
    \frac{n_{\Delta l_\alpha}}{n_{l_\alpha}^\eq}
    +\frac{n_{\Delta H}}{n_H^\eq}
    \Bigg) + \cdots
,
\eal
or equivalently,
\bal
&\frac{d}{d\ln T}\left( \frac{\mu_{l_\alpha}}{T} \right)
=
\gamma^{\rm ID}_{Y_N,\alpha}
\left( 
    \frac{\mu_{l_\alpha}}{T}+\frac{\mu_H}{T}
\right)
 +\cdots
\eal
where
\bal
&\gamma^{\rm ID}_{Y_N,\alpha}
=
\frac{n_N^\eq}{n_{l_\alpha}^\eq}
\frac{K_1(z)}{K_2(z)}
\Gamma_{N\to l_\alpha H},
\eal
and $z=M_N/T$.
Notice that the decay terms do not appear since they were canceled out when we take $\dot n_{l_\alpha} - \dot n_{\bar l_\alpha}$.
$\dot \theta$ dependence enters with the replacement of $\mu_{l_\alpha} \to \mu_{l_\alpha} -\dot \theta/2$. 

\subsection{Complete Boltzmann equations}
The collision terms for the other SM interactions can be easily derived (see, e.g., Ref.\,\cite{Domcke:2020kcp}).
When there is a nonzero background motion of the majoron, the Hamiltonian density in the density matrix will be modified as ${\cal H} \to {\cal H} -\frac{1}{2} \dot \theta J_{B-L}^0$ (see also section.\,\ref{App:dispersion}), so we can effectively replace $\mu_i \to \mu_i + \frac{1}{2}(B-L)_i \dot \theta$ for the SM fermions.

Including the Majorana properties discussed above, the complete Boltzmann equations are
\begin{widetext}
\bal
6H \frac{d}{dx}\hmu_{q_i} &= 
    \gamma_{Y_{u_i}}(\hmu_{q_i}+\hmu_{u^c_i} +\hmu_H) 
    +\gamma_{Y_{d_i}}(\hmu_{d^c_i}+\hmu_{q_i} - \hmu_H)
    +3 \gamma_{\rm WS} \sum_j (\hmu_{l_j}+3\hmu_{q_j})  
    +2 \gamma_{\rm SS} \sum_j (2\hmu_{q_i}+\hmu_{u^c_i}+\hmu_{d^c_i}  )  
\\ 
3H \frac{d}{dx}\hmu_{u^c_i} &= 
    \gamma_{Y_{u_i}}(\hmu_{q_i}+\hmu_{u^c_i} +\hmu_H) 
    + \gamma_{\rm SS} \sum_j (2\hmu_{q_i}+\hmu_{u^c_i}+\hmu_{d^c_i} ) 
\\
3H \frac{d}{dx}\hmu_{d^c_i} &= 
    \gamma_{Y_{d_i}}(\hmu_{d^c_i}+\hmu_{q_i}-\hmu_H)
    + \gamma_{\rm SS} \sum_j (2\hmu_{q_i}+\hmu_{u^c_i}+\hmu_{d^c_i} ) 
\\
2H \frac{d}{dx}\hmu_{l_i} &= 
    \gamma_{Y_{e_i}}(\hmu_{e^c_i}+\hmu_{l_i}-\hmu_H)
    +\gamma^{\rm ID}_{Y_{N,i}}\left( \hmu_{l_i} +\hmu_H -\frac{\dot \theta}{2T} \right) 
    +\gamma_{\rm WS} \sum_j (\hmu_{l_j}+3\hmu_{q_j}) 
\\
H \frac{d}{dx}\hmu_{e^c_i} &= 
    \gamma_{Y_{e_i}}(\hmu_{e^c_i}+\hmu_{l_i}-\hmu_H)
\\
4H \frac{d}{dx}\hmu_{H} &= 
    \gamma_{Y_{u_i}}(\hmu_{q_i}+\hmu_{u^c_i} +\hmu_H) 
    +\gamma_{Y_{d_i}}(-\hmu_{d^c_i}-\hmu_{q_i} + \hmu_H)
    +\gamma_{Y_{e_i}}(-\hmu_{e^c_i}-\hmu_{l_i}+\hmu_H)
    +\gamma^{\rm ID}_{Y_{N,i}}\left( \hmu_{l_i} +\hmu_H -\frac{\dot \theta}{2T} \right) 
\eal
\end{widetext}
where $\hmu \equiv \mu_i/T$ and   $x \equiv \ln T$.
The relaxation rates $\gamma_\alpha$ for the SM Yukawa interactions are well-summarized in Ref.\,\cite{Domcke:2020kcp}.

\section{Equilibrium values}
\label{App:Equilibrium}
The equilibrium values of $\hmu_i$ can be found by solving $d \hmu_i/dx =0$.
When the relaxation rate $\gamma_\alpha>H$, we can impose equilibration condition $\sum_j c_j^\alpha \hmu_j =0$.
These conditions can be explicitly written as
\bal
    \gamma_{Y_{u_i}}:& ~~ \hmu_{q_i}+\hmu_{u^c_i} +\hmu_H=0 \\
    \gamma_{Y_{d_i}}:& ~~ \hmu_{d^c_i}+\hmu_{q_i}- \hmu_H=0 \\
    \gamma_{Y_{e_i}}:& ~~ \hmu_{e^c_i}+\hmu_{l_i}-\hmu_H=0 \\
    \gamma_{\rm WS}:&~~ \sum_j (\hmu_{l_j}+3\hmu_{q_j})  =0 \\
    \gamma_{\rm SS}:&~~ \sum_j (2\hmu_{q_i}+\hmu_{u^c_i}+\hmu_{d^c_i}) =0 
\eal
For interactions with $\gamma_\alpha < H$, we can neglect the corresponding term in the Boltzmann equation, and therefore, we do not impose the equilibration condition for that interaction.  
We assume $\gamma^{D}_{Y_{N_1,i}}$ are always greater than the Hubble rate since we are investigating the scenario around $T\simeq M_N$ in the strong wash-out regime. 

\begin{figure}[t] 
\begin{center}
\includegraphics[width=0.45\textwidth]{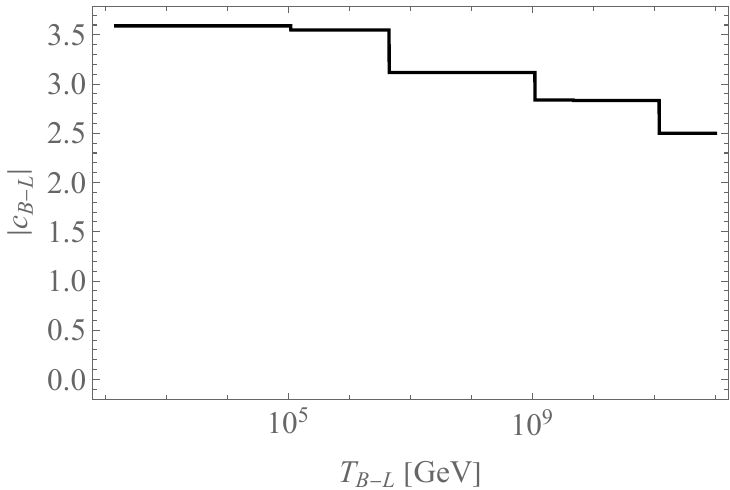} 
\end{center}
\caption{The absolute value of $c_{B-L}(T_{B-L})$ as a function of $T_{B-L}$.}
\label{Fig:cBL}
\end{figure}

We also impose the (hyper) charge neutrality: 
\bal
0=\sum_i (\frac{1}{6} 6 \mu_{q_i}-\frac{2}{3} 3 \mu_{u^c_i}+ \frac{1}{3} 3\mu_{d^c_i}-\frac{1}{2} 2 \mu_{l_i}+\mu_{e^c_i}) + \frac{1}{2}2\cdot 2 \mu_H .
\eal
In addition, there are more conserved numbers depending on the temperature range. Considering all the effects, one can obtain the baryon and lepton asymmetries depending on the temperature region as follows (see Fig.\,\ref{Fig:cBL} for the summary of our estimation).

\begin{itemize}
\item
$T<10^5\,\GeV$: All the interactions are in the thermal bath, and we obtain the resulting $B$, $L$ and $B-L$ asymmetries as follows.
\bal
c_B \simeq -\frac{28}{22} , \quad
c_L \simeq \frac{51}{22}, \quad
c_{B-L}  \simeq -\frac{79}{22}
\eal
\item
$1.1\times 10^5<T<4.5\times 10^6\,\GeV$: $\gamma_{Y_{e_1}}$ (and $\gamma_{Y_{u_1}}$ for $T>10^6\,\GeV$) is decoupled. With imposing $\mu_{e_1^c}=0$, we obtain
\bal
c_B \simeq -\frac{13}{10} , \quad
c_L \simeq \frac{9}{4}, \quad
c_{B-L}  \simeq -\frac{71}{20}
\eal
\item
$4.5\times 10^6<T<1.1\times10^9\,\GeV$: $\gamma_{Y_{d_1}}$ is additionally decoupled. With imposing $\mu_{e^c_1}=0$ and $\mu_{u^c_1}=\mu_{d^c_1}$, we obtain
\bal
c_B \simeq -\frac{20}{17} , \quad
c_L \simeq \frac{33}{17}, \quad
c_{B-L}  \simeq -\frac{53}{17}
\eal
\item
$1.1\times10^9<T<4.7\times 10^{9}\,\GeV$: $\gamma_{Y_{d_2}}$ is additionally decoupled. With imposing $\mu_{e^c_1}=0$, $\mu_{u^c_1}=\mu_{d^c_1}=\mu_{d^c_2}$, and $B_1-B_2=0$, we obtain
\bal
c_B \simeq -\frac{34}{31} , \quad
c_L \simeq \frac{54}{31}, \quad
c_{B-L}  \simeq -\frac{88}{31}
\eal
\item
$4.7\times 10^{9}<T<1.2\times10^{11}\,\GeV$: $\gamma_{Y_{e_2}}$ is additionally decoupled. With imposing $\mu_{e^c_1}=\mu_{e^c_2}=0$, $\mu_{u^c_1}=\mu_{d^c_1}=\mu_{d^c_2}$, and $B_1-B_2=0$, we obtain
\bal
c_B \simeq -\frac{10}{9} , \quad
c_L \simeq \frac{31}{18}, \quad
c_{B-L}  \simeq -\frac{17}{6}
\eal
\item
$1.2\times 10^{11}<T<1.5\times 10^{12}\,\GeV$: $\gamma_{Y_{u_2}}$ (and $\gamma_{Y_{e_3}}$ for $T>1.3\times 10^{12}\,\GeV$) is additionally decoupled. With imposing $\mu_{e^c_1}=\mu_{e^c_2}(=\mu_{e^c_3})=0$, $\mu_{u^c_1}=\mu_{u^c_2}=\mu_{d^c_1}=\mu_{d^c_2}$, and $B_1-B_2=0$, we obtain
\bal
c_B \simeq -1 , \quad
c_L \simeq \frac{3}{2}, \quad
c_{B-L}  \simeq -\frac{5}{2}
\eal
\item
$1.5\times 10^{12}<T<2.5 \times 10^{12}\,\GeV$: $\gamma_{Y_{d_3}}$ is additionally decoupled. With imposing $\mu_{e^c_1}=\mu_{e^c_2}=\mu_{e^c_3}=0$, $\mu_{u^c_1}=\mu_{u^c_2}=\mu_{d^c_1}=\mu_{d^c_2}=\mu_{d^c_3}$, and $B_1=B_2=B_3$, we obtain
\bal
c_B \simeq -\frac{23}{29} , \quad
c_L \simeq \frac{69}{58}, \quad
c_{B-L}  \simeq -\frac{115}{58}
\eal
\item
$2.5\times 10^{12}<T<6\times 10^{12}\,\GeV$: $\gamma_{\rm WS}$ is additionally decoupled. With imposing $\mu_{e^c_1}=\mu_{e^c_2}=\mu_{e^c_3}=0$, $\mu_{u^c_1}=\mu_{u^c_2}=\mu_{d^c_1}=\mu_{d^c_2}=\mu_{d^c_3}$, and $B_1=B_2=B_3=0$, we obtain
\bal
c_B \simeq 0 , \quad
c_L \simeq \frac{69}{44}, \quad
c_{B-L}  \simeq -\frac{69}{44}
\eal
\end{itemize}

\section{Majoron to photon-photon coupling}
\label{App:photonphoton}
For light majoron as in our kinetic misalignment scenario, one may hope that the photon-photon coupling induced by quantum corrections may have a phenomenological signature.
However, that is not the case as we show in the following.
Since the $B-L$ symmetry is anomaly-free, the majoron couplings to gauge bosons involve additional derivatives, e.g. $\partial^2 a F^{\mu\nu} \tilde F_{\mu\nu}$.
The photon-photon interaction is generated at two-loop level\,\cite{Heeck:2019guh}, and the partial decay rate is given as
\bal
\Gamma_{J\to \gamma \gamma} = \frac{|g_{J\gamma}^{\rm eff}|^2}{64\pi} m_J^3,
\eal
where, for $m_J \ll \MeV$,
\bal
g_{J\gamma}^{\rm eff} &\simeq 
\frac{\alpha_{\rm EM}}{16\pi^3 f} \left( \frac{m_J}{\MeV} \right)^2
\Big[
    -0.15\, {\rm tr}[Y_N Y_N^\dagger]
   +0.32 \, (Y_N Y_N^\dagger)_{ee} \nn \\ 
   &+7.5\times 10^{-6} \, (Y_N Y_N^\dagger) _{\mu\mu}
   +2.6\times 10^{-8} \, (Y_N Y_N^\dagger) _{\tau\tau}
\Big].
\eal
To derive an aggressive estimation of phenomenological constraints, we choose the largest $g_{J\gamma}^{\rm eff}$ that is possible along the flavor structure of $Y_N$.
First of all, we use $(Y_N Y_N^\dagger)_{ll}<{\rm tr}[Y_N Y_N^\dagger]$ for $l=e$, $\mu,$ and $\tau$
and also ${\rm tr}[Y_N Y_N^\dagger] \simeq {\rm tr}[Y_N Y_N^{\rm T}]$ so that we obtain
\bal
{\rm tr}[Y_N Y_N^{\rm T}] < 
\frac{2f_J}{v_h^2}
\tilde m_\nu,
\eal
where we also assumed $M_1<M_2<M_3<f_J$ which is true if $N_I$ interactions are perturbative.
Then, the upper bound of $g_{a\gamma}^{\rm eff}$ becomes
\bal
g_{a\gamma}^{\rm eff} < 
g_{a\gamma}^{\rm max} &\simeq
\frac{\alpha_{\rm EM}}{16\pi^3v _h^2} \left( \frac{m_J}{\MeV} \right)^2
\Big[
    0.34\, \tilde m_\nu
\Big] 
\\
&\simeq 4 \times 10^{-21} {\GeV}^{-1} 
\left(\frac{m_J}{\MeV} \right)^2
\left( \frac{\tilde m_\nu}{0.05\,\eV}\right). \nonumber
\eal
Noting that the $X$-ray constraint on an axion-like particle at $m_a\sim \keV$ is roughly $g_{a\gamma} \lsim 10^{-17}\,\GeV^{-1}$~\cite{Foster:2021ngm}, we conclude that it is highly challenging to give constraints on majoron by using the photon-photon interaction.

\end{appendix}

\end{document}